\documentclass[journal,12pt,onecolumn,final]{IEEEtran}
\usepackage{import}
\usepackage{amsmath,amssymb}
\usepackage{mathtools}
\mathtoolsset{showonlyrefs,showmanualtags}
\usepackage{enumitem}
\usepackage{cancel}
\usepackage{graphicx}
\usepackage{bbm}
\usepackage[normalem]{ulem}
\usepackage{hyperref}
\usepackage{amsthm}
\usepackage[ruled]{algorithm2e}
\usepackage{multirow,array}

\usepackage{siunitx}

\usepackage{standalone}
\usepackage{pgfplots}
\usepackage{tikz}
\usepackage{subcaption}

\usepackage{import}
\usepackage{amsmath,amssymb}
\usepackage{mathtools}
\usepackage{etoolbox}

\makeatletter
\let\pragma@iinput=\@iinput
\def\@iinput#1{\xdef\@pragmafile{#1}\pragma@iinput{#1}}
\def\@pragmafile{default}
\def\pragmaonce{%
	\csname pragma@\@pragmafile\endcsname
	\global\expandafter\let \csname pragma@\@pragmafile\endcsname =  
}
\makeatother

\DeclareMathOperator*{\argmin}{arg\,min}

\ifdefined\ceil
\else
\DeclarePairedDelimiter\ceil{\lceil}{\rceil}

\fi
\DeclarePairedDelimiterX{\inprd}[2]{\langle}{\rangle}{#1, #2}

\newcommand{\LnrmS}[1]{\left\lVert #1 \right\rVert}
\newcommand{\Lnrm}[1]{\lVert #1 \rVert}

\newcommand{\Ltwo}[1]{\lVert #1 \rVert_2}

\newcommand{\EV}{{\mathbb{E}}}

\newcommand{\Var}{\mathbb{V}}
\newcommand{\Rel}{\mathbb{R}}

\newcommand{\iid}{i.i.d.}
\newcommand{\defn}{\mathrel{\overset{\makebox[0pt]{\mbox{\normalfont\tiny\sffamily def}}}{=}}}

\newcommand{\cc}[1]{\mathcal{#1}}
\newcommand{\calX}{\cc{X}}

\newcommand{\calV}{\cc{V}}
\newcommand{\calE}{\cc{E}}

\newcommand{\calG}{\cc{G}}

\newcommand{\calN}{\cc{N}}

\newcommand{\emphtwo}[1]{{\textit{#1}}}
\newcommand{\emphone}[1]{{\textbf{#1}}}

\definecolor{darkgreen}{rgb}{0.0, 0.5, 0.0}

\newcommand{\stkout}[1]{\ifmmode\text{\sout{\ensuremath{#1}}}\else\sout{#1}\fi}

\newcommand{\fig}[1]{{Fig.~\ref{fig:#1}}}
\newcommand{\tbl}[1]{{Table~\ref{tbl:#1}}}
\newcommand{\secn}[1]{{Sec.~\ref{secn:#1}}}

\newcommand{\thrm}[1]{{Theorem~\ref{thrm:#1}}}

\newcommand{\eqn}[1]{{\eqref{eqn:#1}}}

\makeatletter
\newcommand*\ifcounter[1]{%
	\ifcsname c@#1\endcsname
	\expandafter\@firstoftwo
	\else
	\expandafter\@secondoftwo
	\fi
}
\makeatother

\ifcsmacro{theorem}{}{

	\ifcounter{chapter}{
		
	}{
		
	}
	
	\newtheorem{theorem}{Theorem}
	
	\newcounter{example}[section]
	
	\newcounter{problem}[section]
	
}

\def\enumtheoremstart{\begin{enumerate}[noitemsep,label=(\roman*)]}
	\def\enumtheoremend{\end{enumerate}}

\usepackage{transparent}

\newif\ifshowcomments
\newif\ifshowdeleted

\newcommand{\devnull}[1]{}
\newcommand{\delete}[1]{{\ifshowdeleted{{\transparent{0.3}#1}}\else\unskip\ignorespaces\fi}}
\newcommand{\comment}[2][]{\ifshowcomments{\printcomment{#1}{#2}}\else\ignorespaces\fi}
\newcommand{\printcomment}[2]{\incolor{#1}{[[\ifx#1\empty\else#1: \fi#2]]}}

\newcommand{\ifequals}[3]{\ifthenelse{\equal{#1}{#2}}{#3}{}}
\newcommand{\case}[2]{#1 #2} 
\newenvironment{switch}[1]{\renewcommand{\case}{\ifequals{#1}}}{}

\definecolor{darkred}{rgb}{0.8, 0.01, 0.1}
\definecolor{darkgreen}{rgb}{0.0, 0.5, 0.0}
\definecolor{darkorange}{rgb}{0.93, 0.35, 0.1}

\newcommand{\incolor}[2]{\ignorespaces
	\begin{switch}{#1}\ignorespaces
		\case{TA}{\color{darkred}}\ignorespaces
		\case{SD}{\color{darkorange}}\ignorespaces
		\case{SK}{\color{darkgreen}}\ignorespaces
		\case{}{\color{red}}\ignorespaces
		#2
	\end{switch}
}

\title{Decentralized optimization with non-identical sampling in presence of stragglers
\author{Tharindu~Adikari 
and~Stark~Draper
\thanks{
This work was supported by Huawei Technologies through a joint project with University of Toronto, 
and the Natural Science and Engineering Research Council (NSERC) of Canada through a Discovery Research Grant. 
}
\thanks{
This paper was presented in part at the IEEE International Conference on Acoustics, Speech and Signal Processing (ICASSP), Barcelona, Spain, May 2020. (\cite{adikari2020decentralized}, \url{https://ieeexplore.ieee.org/document/9053329})
}
\thanks{
The code used for the numerical experiments in this paper is available at \url{https://github.com/thadikari/consensus}.
}
\thanks{
The authors are with the Electrical and Computer Engineering Department, University of Toronto, Toronto, ON M5S 2E4, Canada (e-mail: tharindu.adikari@mail.utoronto.ca; stark.draper@utoronto.ca).
}
}}

\begin{document}
\maketitle
\begin{abstract}
We consider decentralized consensus optimization when workers sample data from non-identical distributions and perform variable amounts of work due to slow nodes known as stragglers. The problem of non-identical distributions and the problem of variable amount of work have been previously studied separately. In our work we analyze them together under a unified system model. We study the convergence of the optimization algorithm when combining worker outputs under two heuristic methods: (1) weighting equally, and (2) weighting by the amount of work completed by each. We prove convergence of the two methods under perfect consensus, assuming straggler statistics are independent and identical across all workers for all iterations. Our numerical results show that under approximate consensus the second method outperforms the first method for both convex and non-convex objective functions. We make use of the theory on minimum variance unbiased estimator (MVUE) to evaluate the existence of an optimal method for combining worker outputs. While we conclude that neither of the two heuristic methods are optimal, we also show that an optimal method does not exist.
\end{abstract}


\pragmaonce  

\newcommand{\mtP}{{P}}
\newcommand{\mtG}{{G}}
\newcommand{\mtW}{{W}}
\newcommand{\mtWh}{\hat{\mtW}}

\newcommand{\numConsen}{{m}}
\newcommand{\graph}{{\calG}}
\newcommand{\vertices}{{\calV}}
\newcommand{\edges}{{\calE}}
\newcommand{\neighs}{{\calN}}

\newcommand{\avgloss}{\bar{\gloss}}
\newcommand{\estgloss}{\hat{\gloss}}

\newcommand{\vvw}{{w}}
\newcommand{\vvwu}[1]{\vvw^{#1}}
\newcommand{\vvwd}[1]{\vvw_{#1}}
\newcommand{\vvwud}[2]{\vvw_{#2}^{#1}}
\newcommand{\vvwji}{{\vvwud{j}{i}}}
\newcommand{\vvwki}{{\vvwud{k}{i}}}
\newcommand{\vvwkj}{{\vvwud{k}{j}}}

\newcommand{\gloss}{g}
\newcommand{\loss}{f}
\newcommand{\lossn}[1]{f_{\text{#1}}}
\newcommand{\lossSum}{{\frac{1}{2}(\loss_1(\vvw)+\loss_2(\vvw))}}

\newcommand{\sumwi}{{\sum}}
\newcommand{\sumworkersl}[1]{{\sum_{#1=1}^{\numDists}}}
\newcommand{\sumworkers}{{\sum_{i=1}^{\numDists}}}
\newcommand{\stepsize}{\eta}

\newcommand{\rvX}{{X}}
\newcommand{\rvXd}[1]{{\rvX_{#1}}}
\newcommand{\rvXi}{{\rvX_i}}
\newcommand{\rvXj}{{\rvX^{j}}}
\newcommand{\rvXij}{{\rvX_{i}^{j}}}

\newcommand{\distQ}{{Q}}
\newcommand{\distQi}{\distQ_{i}}
\newcommand{\numDists}{{n}}
\newcommand{\dimw}{{d}}
\newcommand{\batchSize}{{B}}
\newcommand{\dataSize}{{m}}

\newcommand{\wOptGD}{\vvwu{*}}
\newcommand{\wOptGDd}[1]{\vvwud{*}{#1}}
\newcommand{\wOptSync}{\vvwu{\text{sync}}}

\newcommand{\Eft}{{\EV_{\rvX\sim \distQ}[\loss(\vvwu{t}, \rvX)]}}
\newcommand{\Ef}{{\EV_{\rvX\sim \distQ}[\loss(\vvw, \rvX)]}}
\newcommand{\EFunc}{F}
\newcommand{\EF}{{\EFunc(\vvw)}}
\newcommand{\EFi}{{\EFunc_i(\vvw)}}
\newcommand{\EFj}{{\EFunc_j(\vvw)}}
\newcommand{\EFiNOw}{{\EFunc_i}}
\newcommand{\Efit}{{\EV_{\rvXi\sim \distQi}[\loss(\vvwu{t}, \rvXi)]}}
\newcommand{\Efi}{{\EV_{\rvX\sim \distQi}[\loss(\vvw, \rvX)]}}
\newcommand{\gEFw}{{\nabla\EF}}
\newcommand{\gEF}{{\nabla}}
\newcommand{\gEFiw}{{\nabla\EFi}}
\newcommand{\gEFjw}{{\nabla\EFj}}
\newcommand{\gEFi}{{\nabla_i}}
\newcommand{\gEFj}{{\nabla_j}}
\newcommand{\Egt}{\EV_{\rvX\sim \distQ}[\gloss(\vvwu{t}, \rvX)]}
\newcommand{\Eg}{\EV_{\rvX\sim \distQ}[\gloss(\vvw, \rvX)]}
\newcommand{\EG}{G(\vvw)}
\newcommand{\EGTheta}[1]{G(#1)}
\newcommand{\EGi}{G_i(\vvw)}
\newcommand{\EGj}{G_j(\vvw)}
\newcommand{\EGiTheta}[1]{G_i(#1)}
\newcommand{\Egn}[1]{{\EV_{\rvXd{#1}\sim \distQ_{#1}}[\gloss(\vvw, \rvXd{#1})]}}
\newcommand{\Egi}{\Egn{i}}
\newcommand{\Egit}{{\EV_{\rvXi\sim \distQi}[\gloss(\vvwu{t}, \rvXi)]}}
\newcommand{\EgiTheta}[1]{{\EV_{\rvXi\sim \distQi}[\gloss(#1, \rvXi)]}}
\newcommand{\EgiThetaji}{\EgiTheta{\vvwji}}

\newcommand{\mtGi}[1]{G_{#1}}
\newcommand{\mtWi}[1]{W_{#1}}
\newcommand{\vvbx}{{\mathbf{x}}}
\newcommand{\lpzconst}{L}

\newcommand{\avglosse}{\avgloss_{\text{e}}}
\newcommand{\avglossp}{\avgloss_{\text{p}}}
\newcommand{\varglosse}{\sigma^2_{\text{e}}}
\newcommand{\varglossp}{\sigma^2_{\text{p}}}

\newcommand{\vvx}{{\mathbf{x}}}
\newcommand{\dPdT}{\frac{\partial \ln p}{\partial \theta}}
\newcommand{\dtPdTt}{\frac{\partial^2 \ln p}{\partial \theta^2}}
\newcommand{\dAdT}{\frac{\partial A}{\partial \theta}}
\newcommand{\dBdT}{\frac{\partial B}{\partial \theta}}
\newcommand{\dtAdTt}{\frac{\partial^2 A}{\partial \theta^2}}
\newcommand{\dtBdTt}{\frac{\partial^2 B}{\partial \theta^2}}
\newcommand{\dTdA}{\frac{\partial \theta}{\partial A}}
\newcommand{\dTdB}{\frac{\partial \theta}{\partial B}}

\section{Introduction} \label{secn:intro}
\begin{figure}
\centering
\begin{tikzpicture}
[scale=.3,auto=left,every node/.style={circle,fill=blue!20}]

\node (n0) at (6,7)   {0};     
\node (n1) at (10,10) {1};     
\node (n2) at (20,10) {2};     
\node (n3) at (20,5)  {3};     
\node (n4) at (14,1)  {4};     
\node (n5) at (6,1)   {5};     
\node (n6) at (10,5)  {6};     
\node (n7) at (14,5)  {7};     
\node (n8) at (2,1)   {8};     
\node (n9) at (2,10)  {9};     

\foreach \from/\to in {n9/n8,n9/n5,n9/n1,n9/n0,n8/n5,n7/n6,n7/n3,n7/n1,n6/n1,n5/n0,n5/n3,n5/n4,n3/n1,n3/n2,n2/n1,n0/n1}
\draw (\from) -- (\to);

\end{tikzpicture}
\caption{Workers and the worker connections in a decentralized system. The topology is same as that in Figure 2 in~\cite{ferdinand2018anytime}.}
\label{fig:icasspGraph}
\end{figure}
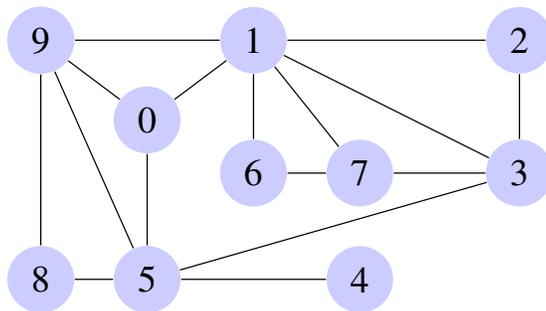

The general system model for decentralized consensus optimization assumes a cluster of workers connected through a network. 
An important property is that there may not be a central coordinator amongst workers (unlike in master-worker systems). 
\fig{icasspGraph} presents an example of how $10$ workers may be connected in a decentralized manner. 
Each worker is associated with a utility function 
and the goal is collaboratively and iteratively to optimize the sum of utility functions using only local computations. In each iteration, workers first perform local computations and then synchronize their results through some consensus mechanism. 
We refer to the two phases as `compute' and `consensus'. 
Workers can arrive either at perfect or approximate consensus depending on the consensus scheme used. 
For example the analysis in \cite{dekel2012optimal} assumes perfect consensus which can be achieved 
using 
operations in the Message Passing Interface (MPI) standard. However, recently there has been significant interest in approximate averaging 
methods due to a number of attractive features. Such features include that these schemes are simple to implement, they support asynchronous operation, and they work well with dynamic network structures \cite{duchi2011dual, tsianos2016efficient}. 

A canonical application of decentralized consensus optimization is distributed machine learning. With large datasets, it is desirable to assign smaller subsets of the data to each worker, with workers collaborating to find an optimal model for the entire dataset. In this paper we study the global convergence of such a system that employs stochastic gradient descent (SGD) when data distributions at workers are non-identical \emph{and} when workers perform variable amounts of work per iteration. We summarize the contributions of this paper with its outline as follows. In \secn{relatedwork} we discuss the related work in this theme of studies. In \secn{background} we outline the system model that we work with. 
Since workers perform variable amounts of work, the level of accuracy in worker outputs vary from one worker to the other. The worker outputs must be combined in a way that leads to the fastest convergence of the system. In \secn{heuristicsec} we discuss two heuristic methods of combining the worker outputs. We numerically and theoretically analyze the performances of the two methods and prove their convergence with SGD. 
Subsequently, in \secn{icasspoptimality} we discuss the existence of an optimal method of combining the worker outputs.

\section{Related work} \label{secn:relatedwork}
There has been a significant amount of recent interest in the application of decentralized optimization to machine learning. 
A variety of system models have been considered \cite{dekel2012optimal, duchi2011dual, tsianos2016efficient, ferdinand2018anytime}. 
These studies differ in their assumptions of the network topology, the data model, the averaging method, and the underlying optimization algorithm. 
For example, the scheme in \cite{dekel2012optimal} employs exact averaging 
and stochastic gradient descent. 
In contrast, \cite{duchi2011dual, tsianos2016efficient, ferdinand2018anytime} rely on random walk-based approximate consensus and dual averaging \cite{nesterov2009primal, xiao2010dual}. 
The authors of \cite{ferdinand2018anytime} consider the optimization problem in the presence of stragglers (slow workers) and introduce Anytime MiniBatch (AMB) \cite{ferdinand2017anytime} to exploit stragglers. The idea behind AMB is to allocate all workers a fixed amount of time for gradient computations in each iteration 
so that slow workers do not hold up the system. 
A time limit is imposed on the consensus phase as well to ensure the system does not stall due to random communication delays. Workers then apply the gradients obtained in the consensus phase and proceed to the next iteration. 

The closest system models to ours are those of \cite{duchi2011dual} and \cite{ferdinand2018anytime}. 
We consider a generalized model that captures important aspects of each paper. Specifically, \cite{duchi2011dual} assumes non-identical data distributions at the workers but 
does not take into account that the gradients may be computed using different amounts of data per iteration. 
On the other hand, \cite{ferdinand2018anytime} considers identical data distributions 
but variable amount of gradient computations. 
In this case, gradients from workers are weighted according to the amount of work performed. 
In \secn{heuristicsec} we discuss how these two ideas can be combined to achieve faster convergence when workers complete variable amounts of work and data distributions across workers are different. 

\delete{
++ None of these papers consider the convergence or convergence rate for non-identical distributions and proportionate gradients. 

++ differentiates from ~\cite{dekel2012optimal} and \cite{ferdinand2018anytime} where only the special case of $\distQi=\distQ$ is considered. ++ cat and dogs distributed across graph

++ positioning of this work is as follows.

TABLE:::	
samples for gradient estimate	consensus rounds	distributions across workers	gradient weighting
Duchi	stochastic - 1 	single	different	equal
Rabbat	stochastic - 1	multiple 	same	equal
AMB	variable variable same proportional
Proposed variable multiple different proportional
}

\section{System model} \label{secn:background}
Similar to \cite{dekel2012optimal, duchi2011dual, tsianos2016efficient, ferdinand2018anytime}, we consider a distributed optimization problem defined on an undirected graph $\graph(\vertices, \edges)$. Here, $\vertices$ and $\edges$ stand for the set of vertices and the set of edges in the graph. 
A vertex represents a worker and an edge represents a bi-directional communication link. 
Specifically, $(i,j)\in\edges$ means that there exists a direct link between workers $i$ and $j$. 
Note that communication between any two workers that are not directly connected must be relayed through others. 
We denote the number of workers $|\vertices|$ by $\numDists$. 
Also, we assume that the graph is connected, i.e., there is a path that traverses edges and connects any vertex to any other vertex in the graph. 

\subsection{Worker data distributions}
Workers share a cost 
function $\loss:\Rel^\dimw\times\calX\to\Rel$ 
which is parameterized by a vector $\vvw\in\Rel^\dimw$. The cost for data point $\rvX\in\calX$ is $\loss(\vvw, \rvX)$. 
We assume that $\loss$ is differentiable, $\lpzconst$-Lipschitz convex in $\vvw$ for all $\rvX$ and has bounded gradient, i.e., $\Lnrm{\nabla_\vvw\loss(\vvw, \rvX)}\leq \lpzconst$. 
Also, we assume that the data distributions across workers are non-identical. The $i$th worker can only sample data from distribution $\distQi$. 
We define the expected cost for the $i$th worker with respect to its data distribution as
$\EFi \defn \Efi$. 
Here, the random variable $\rvX$ is an abstraction for common optimization problems such as unsupervised and supervised learning. For example, in the supervised case $\rvX$ represents the (data, label) pair and $\distQi$ is the joint distribution of the pair. 

We are interested in finding a globally optimal parameter vector considering \emph{data across all} workers. 
To this end we define a mixture distribution $\distQ \defn \sumworkers\gamma_i \distQ_i$. 
The priors $\gamma_i\geq0$ represent the relative importance of each distribution. They are assumed known and satisfy $\sumworkers\gamma_i=1$. 
For example, if worker datasets are finite 
the normalized sizes of datasets can be taken to be $\gamma_i$. 
However, in our analysis we consider the more general situation where the size of each dataset is not necessarily proportional to its prior. 
In this case we define the global objective across workers to be 
\begin{align}
\EF \defn \Ef = \sumworkers \gamma_i\Efi = \sumworkers \gamma_i\EFi. \label{eqn:defFw}
\end{align}
We want to design a system that enables 
all workers to converge to 
\[\wOptGD \defn \argmin_{\vvw\in\Rel^\dimw} \EF.\]
Let us denote the minimizer of $\EFi$ by $\wOptGD_i$. 
Note that with no prior assumptions on the $\distQ_i$, $\wOptGD_i \neq \wOptGD_j$ for $i\neq j$. 
In other words, the $i$th worker has access to only $\EFiNOw$ yet its goal is to collaboratively and iteratively find $\wOptGD$. 
In subsequent sections we use 
\[\gloss_i(\vvw, \rvX)\defn\nabla_\vvw\loss_i(\vvw, \rvX)\] 
to denote the gradient computed at the $i$th worker using data point $\rvX$. We note that the bounded gradient assumption on $\loss$ also tells us that $\Lnrm{\gEFiw}\leq \lpzconst$ which can be shown by using Jensen's inequality and the convexity of norms. 

\comment{
In \secn{analysis} we will make use of the result 
$\gEFiw^T\gEFjw
\leq\lpzconst^2$, which can be shown using the Cauchy-Schwarz inequality. 
technically the proofs in analysis section should consider that $\gloss_i$ and $\gloss_j$ are independent only conditionally, i.e., given the history of parameter vectors and the past gradients.}

\subsection{Variable computations} \label{secn:varComp}
Let us now consider a case analogous to \cite{ferdinand2018anytime}, i.e., when stragglers are present. 
In this paper we assume a \emph{homogeneous} cluster. With this assumption straggler statistics can be considered \iid{}~across workers, although the data distributions may differ. 
In any iteration, the $i$th worker computes gradients from $b_i$ data samples. 
Specifically, for a given $\vvw$ the worker computes $\gloss_i(\vvw, \rvX)$ using $b_i$ realizations of $\rvX$ sampled from $\distQi$. 
We denote by $\avgloss_i(\vvw)$ the average of the $b_i$ realizations of $\gloss_i(\vvw, \rvX)$. 
We do not include the iteration index in $b_i$ or $\avgloss_i$ to avoid clutter. 
The $b_i$ themselves are random variables taking values in $\{1,2,\dots\}$. 
We assume that the $b_i$ are \iid{}~for all $i\in[\numDists]$ and across iterations. This means $b_i$ may differ from iteration to iteration even for the same worker. 
Our assumptions on $b_i$ are consistent with homogeneous computing clusters whose straggler statistics are identical across compute nodes. We also note that $b_i\geq1$ and each worker computes at least one gradient sample. 
The latter assumption will prove to be useful in our analysis in \secn{analysis}. 

\delete{
The general template for consensus optimization considered in \cite{duchi2011dual, tsianos2016efficient, ferdinand2018anytime} involves the workers iterating two phases, namely the compute phase and the consensus phase. In the former, workers estimate $\EGi$ using a single realization of $\rvXi\in\distQ_i$ or multiple realizations, corresponding to stochastic gradient descent and mini-batch gradient descent.

The general template for decentralized optimization involves the workers iterating two phases, namely the `compute phase' and the `consensus phase'. In the former, workers separately estimate expected gradients of their utility functions using one data sample of multiple samples, corresponding to stochastic gradient descent and mini-batch gradient descent. In the consensus phase the workers compute the average of estimates using the distributed averaging method. The workers update their parameter vectors 
	
\emphone{Remark 1}: The objective in \eqn{wOptSum} is a sum of convex functions. Absorbing the constant $\gamma_i$ into $\EFi$ precisely gives us the problem considered in~\cite{duchi2011dual}. 
In iteration $t$ let the $i$th worker compute $\gloss(\vvw, \rvXi)$ by sampling $\rvXi\in\distQ_i$. 
In~\cite{duchi2011dual} this quantity is denoted as $g_i(t)$ in equation (5a). 
To get the corresponding consensus update it suffices to use $\gamma_i\gloss(\vvw, \rvXi)$ in place of  $g_i(t)$ in equation (5a). By the convergence proof in \cite{duchi2011dual}, this scheme guarantees worker parameters converge to $\wOptGD$.

Theorem 4 in Duchi is for stochastic gradients. I quote from the theorem: "...at each round of the algorithm agent i receives a vector $\hat{g}_i(t)$ from an oracle...". The quantity $\hat{g}_i(t)$ is the gradient (at iteration $t$ in $i$th worker) corrupted with zero-mean bounded-variance noise. The noise is described in (12) and the gist of it is:
	1. $E[\hat{g}_i(t) | history of gradients, weights at i-th node] = g_i(t)$ (the true gradient),
	2. variances of all $\hat{g}_i(t)$ are bounded by $\lpzconst^2$.

\emphone{Remark 2}: Let us now consider the analogous case to \cite{ferdinand2018anytime}, i.e., when stragglers are present. In this case workers average multiple samples of $\gloss(\vvw, \rvXi)$ to estimate $\EGi$, and also do multiple consensus rounds within separate time limits. This can be though of as a generalization of the scheme in Remark 1, where now the workers have better estimates and better consensus. Simply starting the consensus with estimated gradient weighted by $\gamma_i$ will get our scheme to desired convergence as per the proof in \cite{ferdinand2018anytime}.
}

\subsection{Consensus phase} \label{secn:consensus}
\begin{figure}
	\centering
	\includegraphics[width=0.7\textwidth]{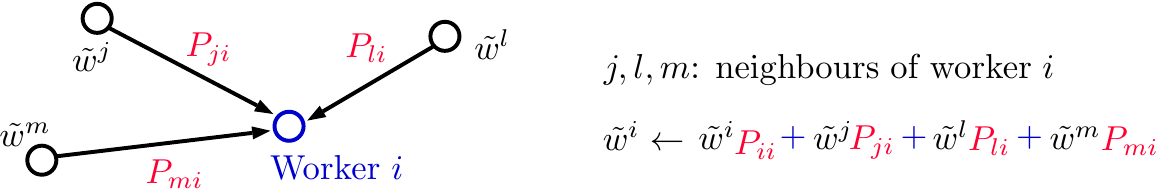}
	\caption{Example of one random walk-based consensus round. 
	The message vectors are denoted by $\tilde{w}$. Worker $i$ computes the weighted sum of $\tilde{w}^i$ and  the messages coming from its neighbours. 
	At the end of one consensus round the $i$th worker updates the message vector as shown.}
	\label{fig:icassp_worker_communication}
\end{figure}

In this paper we consider gradient descent as the core optimization algorithm, whereas \cite{duchi2011dual} and \cite{ferdinand2018anytime} use dual averaging. 
After computing $\avgloss_i$ workers locally apply gradients and move to the consensus phase. 
The goal of this phase is to let workers synchronize by averaging across the local parameter vectors. We show that this strategy drives all workers to converge to $\wOptGD$. 
Let $k\in\{0,1,2,\dots\}$ denote the iterate and let $\vvwki$ denote the parameter vector at the $i$th worker in iteration $k$. 
We assume that 
all workers are initialized to $\vvwud{0}{i}=\vvwu{0}$. 
As per \secn{varComp}, the $i$th worker computes $\avgloss_i(\vvwki)$ in the $k$th iteration. 
In this paper we use the random walk-based approximate consensus proposed in \cite{duchi2011dual}. 
Let $\mtP\in\Rel^{\numDists\times\numDists}$ be a doubly stochastic matrix whose $i,j$th element $\mtP_{i,j}>0$ only if $(i,j)\in\edges$. We denote the $\numConsen$th matrix power of $\mtP$ by $[\mtP]^\numConsen$. 
Methods for generating this type of a matrix include those based on Metropolis-Hastings weights \cite{xiao2007distributed} and graph Laplacians \cite{duchi2011dual}. 
Since $\mtP$ is a doubly stochastic matrix, all entries in $[\mtP]^\numConsen$ converge to $\frac{1}{\numDists}$ as $\numConsen$ grows. 
This can be shown by first observing that the all-ones vector is a left \emph{and} a right eigenvector of $\mtP$, corresponding to the largest eigenvalue $1$. One can then consider an eigendecomposition of $\mtP$ to show convergence. 
The rate of convergence is determined by the second largest eigenvalue of $\mtP$. 
In random walk-based consensus the $i$th worker takes a weighted sum of the message vectors from itself and its neighbours. 
Entries in the $i$th row of $\mtP$ are taken as the weights, and as per the construction of $\mtP$ all non-neighbours have zero weights. 
This process is illustrated in \fig{icassp_worker_communication}. 
All workers receive the average of message vectors if the exchanging and summing operations are iteratively carried out for many rounds. 

\tbl{sysparamstable} summarizes the list of the system parameters and the notations used in the paper. 
Note that the table consists of parameters that will be introduced in next sections as well. 

\begin{table}[]
\renewcommand{\arraystretch}{1.4}
\caption{Definitions of system parameters.}
\label{tbl:sysparamstable}
\centering
\begin{tabular}{|c|c||c|c|}
\hline
\textbf{Parameter}      & \textbf{Definition}								& \textbf{Parameter}    & \textbf{Definition}    			          					\\\hline\hline
$\distQi$ 				& data distribution of $i$th worker				    & $\avglosse$			& $\sumwi\gamma_i\avgloss_i$ 									\\\hline
$\distQ$  				& $\sumworkers\gamma_i \distQ_i$ 				    & $\varglosse$			& $\Var(\avglosse)$												\\\hline
$b_i$	  				& number of gradient samples at $i$th worker	    & $\avglossp$			& $\sumwi \frac{\numDists b_i}{b}\gamma_i\avgloss_i$			\\\hline
$b$		  				& $\sumworkers b_i$ 							    & $B$					& $\{b_1,\dots,b_\numDists\}$									\\\hline
$\gloss_i = \gloss_i(\vvw, \rvX)$	& $\nabla_\vvw\loss_i(\vvw, \rvX)$ 	    & $\mu_1$				& $\EV[b_i/b]$                     								\\\hline
$\avgloss_i = \avgloss_i(\vvw)$ 	& average of $b_i$ realizations of $\gloss_i(\vvw, \rvX)$    & $\mu_2$				& $\EV[{1/}{b_i}]$                 			\\\hline
$\gEFi$					& $\gEFiw$										    & $\mu_3$ 				& $\EV[{b_i/}{b^2}]$											\\\hline
$\gEF$ 					& $\gEFw$										    & $c_i$					& $\frac{b_i}{b}-\mu_1$											\\\hline
$\sumwi$ 				& $\sumworkers$									    & $s^2$					& $\EV[c_i^2]$													\\\hline
\end{tabular}\end{table}

\section{Heuristic gradient estimators} \label{secn:heuristicsec}
Let $\mtWi{k}$ and $\mtGi{k}$ be the $\numDists$-column matrices whose $i$th columns are $\vvwki$ and $\avgloss_i(\vvwki)$ respectively. 
After computing $\avgloss_i(\vvwki)$, workers can apply the gradient and update the parameter vector in any manner that leads to the fastest convergence. 
We discuss in \secn{heuristicsec} two heuristic methods of applying gradients at workers. 
In \secn{twoweighting} we present the two methods, 
in \secn{icasspexperiments} we present numerical results obtained with the two methods, 
and finally in \secn{analysis} we present a convergence analysis of the two methods. 
We show in \secn{analysis} that the method that produces the unbiased estimate of $\gEFw$ with the lower variance leads to a faster convergence. We call the two methods heuristic gradient estimators to reflect that they are estimating $\gEFw$.

\subsection{Two gradient weighting methods} \label{secn:twoweighting}

\subsubsection{Equal weighting}
The idea in this scheme is to locally apply gradient as $\vvwki-t\numDists\gamma_i\avgloss_i(\vvwki)$ and use the result to perform $\numConsen$ random walk-based consensus rounds. 
Here, $t>0$ is the step size. 
Workers then take the output of consensus as $\vvw_j^{k+1}$ and proceed to the next iteration. 
Let $V_1$ be the diagonal matrix whose diagonal is $(\gamma_1,\dots,\gamma_\numDists)$. 
Then $\vvw_j^{k+1}$ is given by the $j$th column of 
\[\mtW^{k+1} = (\mtW^{k}-t\numDists V_1\mtG^k)[\mtP]^\numConsen.\] 
Since $\mtP$ is a doubly stochastic matrix, the product $\numDists[\mtP]^\numConsen$ converges to the all-ones matrix as $\numConsen$ grows. 
In the limit of $\numConsen$, all columns of $\mtW^{k+1}$ converge to 
\begin{equation}\label{eqn:equalweighting1}
\vvw_j^{k+1} =  \frac{1}{\numDists}\sumworkers\vvwki - t\sumworkers\gamma_i\avgloss_i(\vvwki).
\end{equation}
Since all workers are initialized to $\vvwud{0}{i}=\vvwu{0}$, \eqn{equalweighting1} is equivalent to perfect consensus and the parameter vectors are identical across workers for all iterations. 
We can equivalently write the update equation for all workers as 
\begin{align} 
\vvwud{k+1}{j} =  \vvwkj - \stepsize_k\sumworkers\gamma_i\avgloss_i(\vvwki), \label{eqn:equalweighting2}
\end{align} 
irrespective of $j$. 
The term `equal weighting' is used in the sense that $\gamma_i\avgloss_i(\vvwki)$ 
from all workers are treated equally when computing $\vvw_j^{k+1}$. This is in contrast to the scheme we describe next.
\delete{
	Upon hearing from all neighbours, the $i$th worker updates its parameters by summing the 
	neighbour parameters weighted by the non-zero elements in $i$th column of $\mtP$. 
	Concisely, after the first consensus round, the parameter vector at $i$th worker is given by the $i$th column of the matrix $\mtG^1\defn\mtG^0\mtP$. 
	Note that due to the construction constraints of $\mtP$, combining parameters from workers that are not neighbours is not allowed. 
	This parameter exchange is performed $k$ times 
	after which the $i$th worker has the $i$th column of the matrix $\mtG^k\defn\mtG^{k-1}[\mtP]=\mtG^0[\mtP]^k$. 
	
	At this point the consensus phase finishes and $\vvwud{t+1}{i}$ is made equal to the $i$th column of $\mtG^k$, which is used for the compute phase in the next iteration. 
	In practice the workers are not able to compute $\EGiTheta{\vvwud{t}{i}}$  exactly, but are only able to approximate. Usually each worker has a finite dataset sampled from $Q_i$ and the average of the corresponding gradients is considered as $\EGiTheta{\vvwud{t}{i}}$. 
	
	{\emphone{Consensus phase}:} In this step the $i$th worker participates in \emphtwo{$\numConsen_i^t$ consensus rounds}. Specifically, in each consensus round workers exchange the updated parameters with their immediate neighbours.. 
	To achieve consensus we use the following distributed averaging variant. 
	
	The workers iterate these two phases until some convergence criteria is met. We would like to design the matrix $\mtP$ such that for large $k$ 
	\begin{equation} \label{eq:uwrapGDi}
	\vvwud{t+1}{i} = \vvwud{t}{i} - \eta_t\EGTheta{\vvwud{t}{i}} = \vvwud{0}{i}  - \sum_{r=0}^{t}\eta_{r} \EGTheta{\vvwud{r}{i}}
	\end{equation}
	for all $i$. 
	Note that for $\vvwud{0}{i}=\vvwu{0}$, \eqn{uwrapGDi} produces same iterates as~\eqn{updateGD} for all $i$. 
	In other words, all workers follow the same gradient descent path as if it is done on one worker, which guarantees to converge to $\wOptGD$. But there are two main sources of deficiencies in play. It is not possible to compute the expected gradients since distributions are unknown, and in practice $k$ cannot be made arbitrarily large. The latter causes workers to receive slightly different $\vvwud{t+1}{i}$ vectors, which is termed `imperfect consensus' \cite{ferdinand2018anytime}. 
	In this paper we analyse the convergence across workers when these two deficiencies are present, and the worker distributions are not identical. how to design P such that the weighted average is computed?
}

\subsubsection{Proportional weighting}
In the previous scheme, although workers compute $\avgloss_i(\vvwki)$ using different values for $b_i$, the $b_i$ are not taken into account when combining the estimates. 
For example, the $i$th worker may compute $\avgloss_i(\vvwki)$ with $b_i=10$ whereas the $j$th worker may compute $\avgloss_j(\vvwkj)$ with $b_j=100$. The latter would be a less noisy estimate. 
Naturally, one should ask whether we can do better by taking into consideration the confidence of each gradient estimate. This gives rise to the following scheme. 

Let $b\defn \sumworkers b_i$. In this scheme we want to formulate an initial message that enables all workers to receive 
\begin{align} 
\vvwud{k+1}{j} = \vvwkj - \stepsize_k\sumworkers\frac{\numDists b_i}{b}\gamma_i\avgloss_i(\vvwki)
\label{eqn:propweighting}
\end{align} 
in the limit of $\numConsen$. 
Compared to the equal weighting scheme, now $\gamma_i\avgloss_i(\vvwki)$ is weighted by its relative confidence $\frac{\numDists b_i}{b}$. 
In the limit of $\numConsen$, all workers receive the desired parameter vector 
if the initial message of the $i$th worker is set to 
$\vvwki-\numDists^2\frac{b_i}{b}\gamma_i\avgloss_i(\vvwki)$. 
Let $V_2$ be the diagonal matrix whose diagonal is $(\frac{\numDists b_1}{b}\gamma_1,\dots, \frac{\numDists b_\numDists}{b}\gamma_\numDists)$. 
We can obtain the matrix form of approximate consensus by letting 
\[\mtWi{k+1} = (\mtWi{k}-\stepsize_k\numDists V_2\mtGi{k})[\mtP]^\numConsen.\] 

Note that the equal weighting scheme can be recovered by setting a constant for all $b_i$. 
Although equal and proportional schemes are quite similar, the convergence properties of the latter may not be immediately apparent. This is because the $\frac{b_i}{b}$ themselves are random variables, and at a given iteration a larger $b_i$ pulls  $\vvwud{k+1}{j}$ towards $\wOptGD_i$ for any $j$.

Next, in \secn{icasspexperiments} we present a few numerical experiments that demonstrate emprirical performance of equal and proportional weighting schemes with a real dataset. 
In \secn{analysis} we theoretically analyze the convergence properties of the two schemes and we prove that both schemes make all workers converge to $\wOptGD$. 

\comment{The initial message in the proportional weighting scheme consists of $b$. Comment on how does the $i$th worker compute $b$ without having access to all other $b_j$?

In the fully distributed set-up, how to divide by ${b}$ because the workers only know individual ${b_i(t)}$ values.

For the equal and proportional weighting schemes do the workers need access to $n$ and $b$? Can we design an algorithm to achieve global optimization but each worker only having access to local information (itself and the neighbours)? A good example of achieving a global goal with local information is the construction of $P$ using the metropolis weights. Metropolis method only require local connectivity of workers to produce a $P$ matrix in a way that agrees with the connectivity graph. 

To circumvent the issue with $n$ and $b$, one could treat $b$ as a constant and absorb $n$ and $b$ into the learning rate. Therefore effectively not worrying about them at all. But in this case all workers need to know a proper learning rate. Another way of estimating $b$ is to maintain a running average of batch sizes at each worker using its past computations. This estimate can be treated as the $b$ which is then used at the denominator of the message. 

The last case is probably solved by the Push-Sum protocol proposed in~\cite{kempe2003gossip}. 
}

\subsection{Numerical results} \label{secn:icasspexperiments}
\begin{figure}
\centering\includegraphics[width=0.55\textwidth]{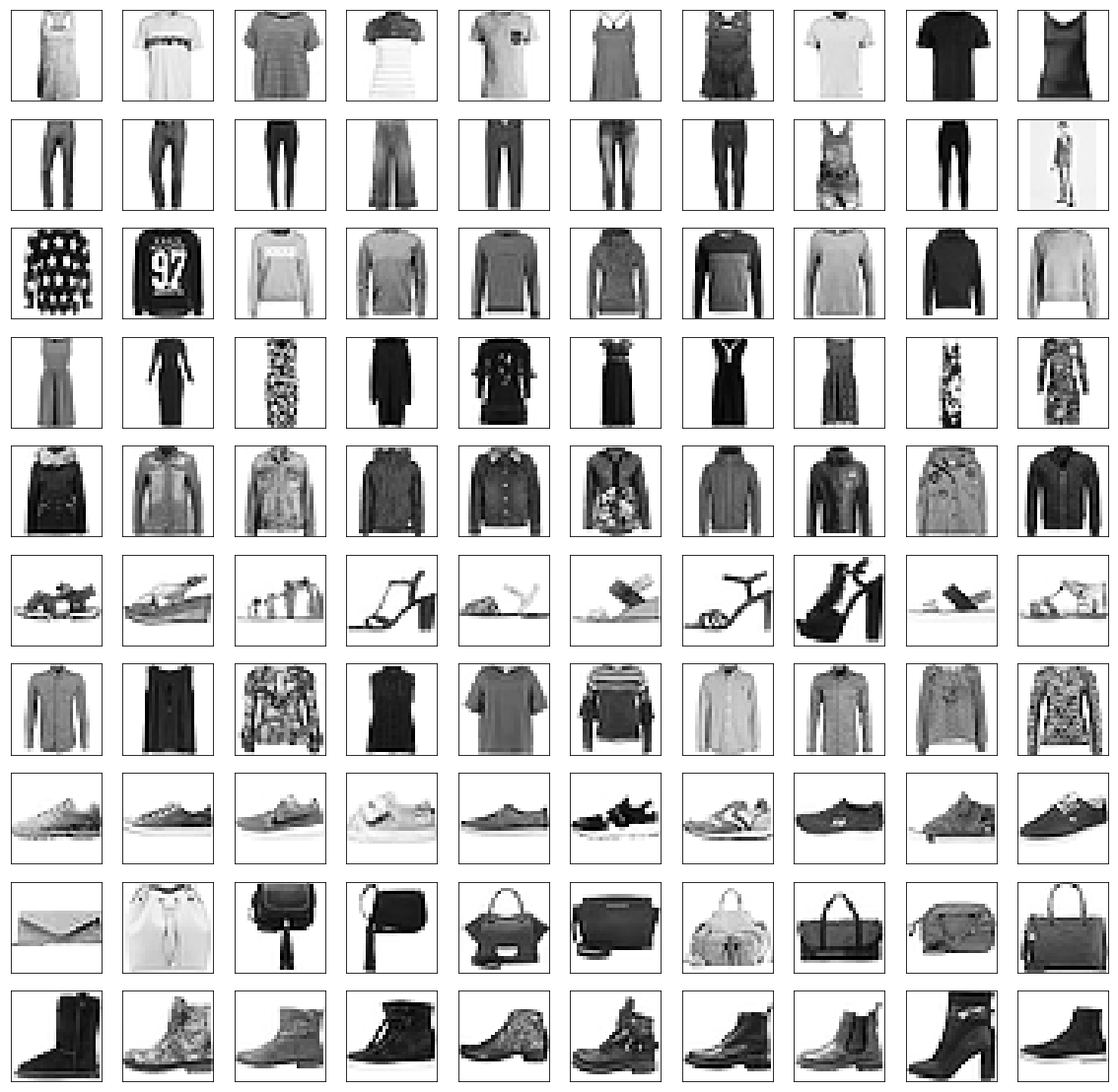}
\caption{A few examples from the 10 classes in the Fashion-MNIST dataset (one class per row).}
\label{fig:fashion_mnist}
\end{figure}

We present experiments performed on $10$ workers using the Fashion-MNIST dataset \cite{xiao2017fashion}. 
As shown in \fig{fashion_mnist}, the dataset consists of $28\times 28$ grayscale images of fashion items that belong to $10$ classes. 
Two sub-datasets that are of sizes $50,000$ and $10,000$ are available for training and testing. 
To simulate different distributions we partition the original dataset into 10 groups and assign each to one of the 10 workers. 
The partitioning is done in a way that only the $i$th worker receives images of class $i$. 
This is to ensure that the distributions of different workers are distinct. 
The workers are connected as per the topology shown in \fig{icasspGraph}. 
The label of each worker in \fig{icasspGraph} indicates $i$, the class index of the images assigned to that worker.

The matrix $\mtP$ is defined using Metropolis-Hastings method and its second largest eigenvalue 
is $0.888$. 
We select cross entropy loss of a multinomial logistic regression classifier as the cost function. The distinction between the two methods considered in this work is best illustrated when the size of the parameter vector $\vvw$ is large. 
To increase the size of $\vvw$ we include one hidden layer in the classifier \emph{without} using an activation function. The resulting cost function, which we denote by $\lossn{lr}$, remains convex in $\vvw$ but the size of the parameter vector is now increased. 
We also test the non-convex function obtained by applying $\max(0,\cdot)$ as the activation in the hidden layer. This cost function is denoted by $\lossn{nn}$.

\begin{figure}
\begin{center}
\begin{minipage}[b]{0.6\linewidth}
\centering
\centerline{\includegraphics[width=\textwidth]{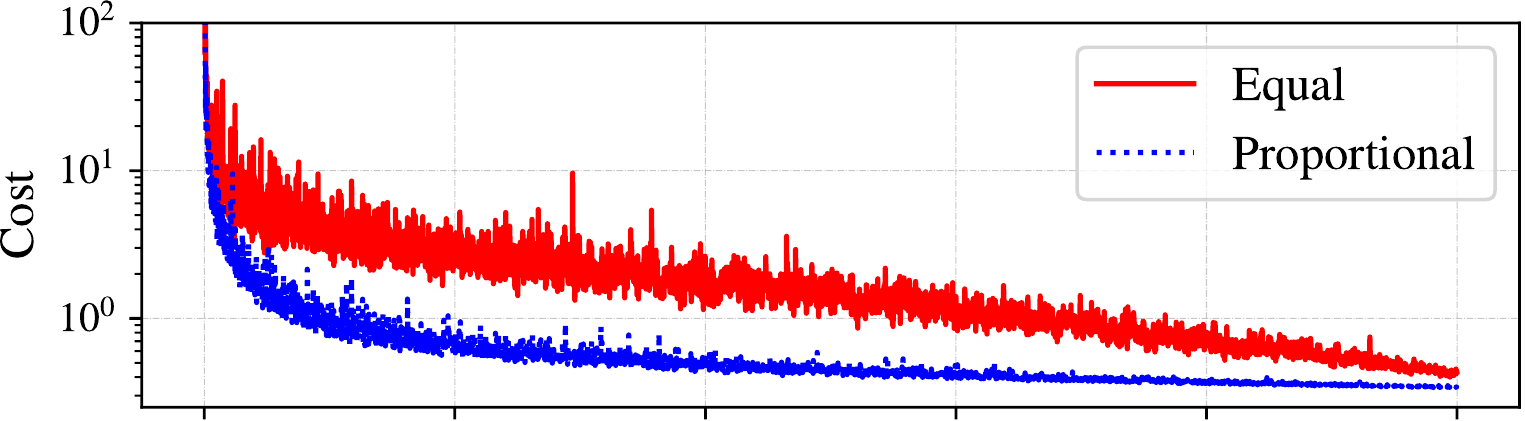}}
\end{minipage}
\begin{minipage}[b]{0.6\linewidth}
\centering
\centerline{\includegraphics[width=\textwidth]{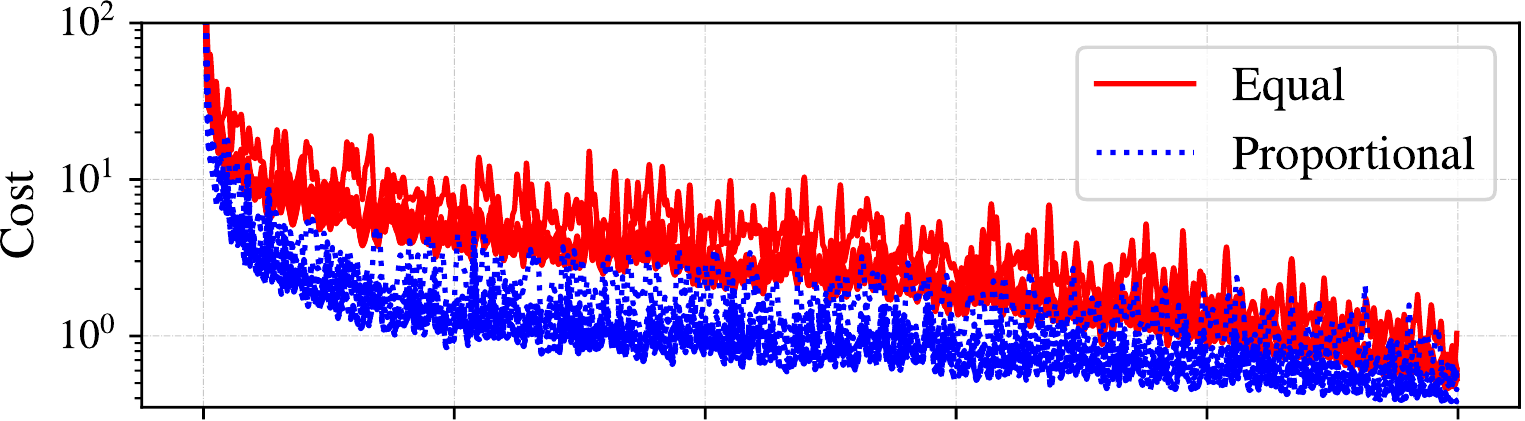}}
\end{minipage}
\begin{minipage}[b]{0.6\linewidth}
\centering
\centerline{\includegraphics[width=\textwidth]{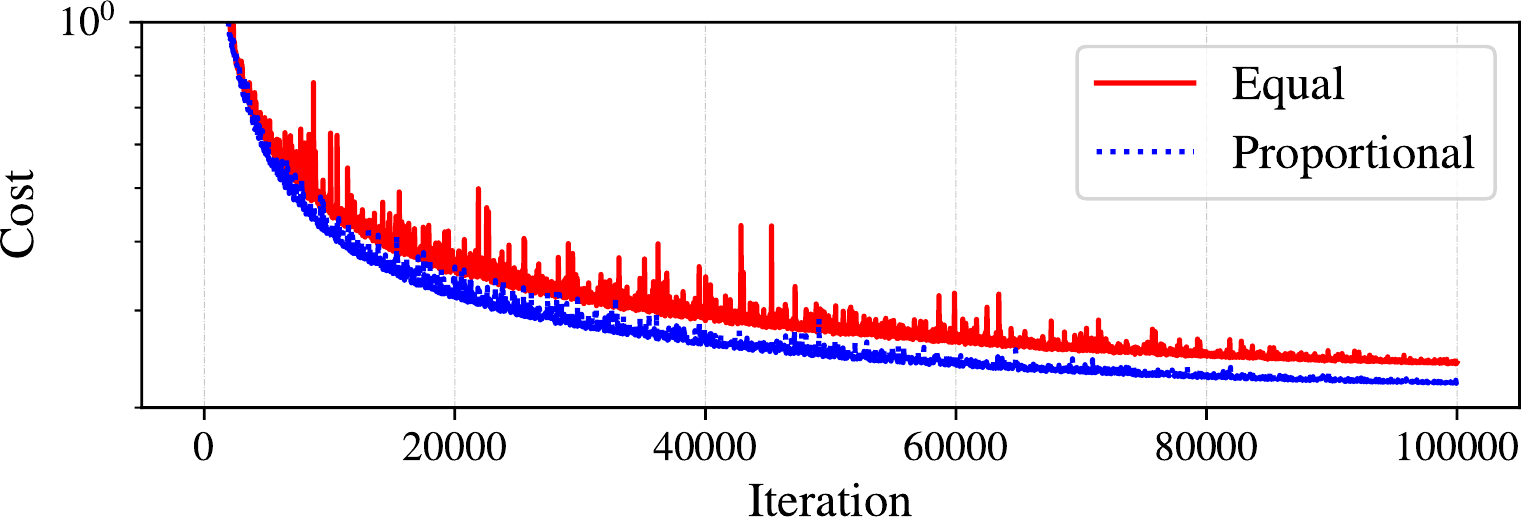}}
\end{minipage}
\caption{Results for $\lossn{lr}$ with perfect consensus (top), for $\lossn{lr}$ with approximate consensus (middle), and for $\lossn{nn}$ with perfect consensus (bottom). 
{To suppress noise, the data in the middle figure have been smoothed using a Gaussian filter with standard deviation $5$.}
}\label{fig:icasspfigures}
\end{center}
\end{figure}

To simulate stragglers we sample $b_i$ from a Bernoulli distribution in an \iid{} manner. 
A worker chooses $b_i=60$ data points with probability $0.8$ and $b_i=1$ with probability $0.2$. 
These choice of numbers make the distribution of $b_i$ heavily skewed and result in quite noisy gradients at some workers. 
Convergence results are shown in \fig{icasspfigures}. 
The costs of all workers are identical in perfect consensus. 
For approximate consensus $\numConsen$ is set to $10$. 
In this case the workers have different weights (and costs) as iterations progress, therefore we plot the costs of all 10 workers. Note that these are the values of $\EFunc(\vvwki)$, i.e., the costs with respect to the global data distribution. 
We observe that the proportional method outperforms equal weighting under both consensus schemes. 
In particular, the equal weighting plots are noticeably more noisy than those of proportional weighting.

\subsection{Convergence analysis} \label{secn:analysis}
In this section we prove the convergence of the two weighting schemes by assuming perfect consensus amongst workers. 
Recall that with perfect consensus at the end of the consensus phase all workers possess the same gradient estimate, and apply on the same parameter vector. 
This means in each iteration workers apply the same gradient and obtain identical parameter vectors. 
To simplify notation we denote $\gloss(\vvw, \rvX)$ for $\rvX\sim\distQ_i$ by $\gloss_i$, and $\avgloss_i(\vvw)$ by $\avgloss_i$. Recall that $\avgloss_i$ is the average of $b_i$ instances of  $\gloss_i$. 
In the interest of reducing clutter, 
we use $\gEFi$, $\gEF$ and $\sumwi$ to denote $\gEFiw$, $\gEFw$ and $\sumworkers$ respectively. 
Before proceeding to the main proof we first state the following few properties that will become useful. 

\begin{enumerate}
\item 
Note that \begin{align}
\EV[\gloss_i] = \EV_{\rvX\sim \distQi}[\gloss(\vvw, \rvX)] = \nabla\EV_{\rvX\sim \distQi}[\loss(\vvw, \rvX)] = \gEFi. \label{eqn:EVgi}
\end{align}
For a given $b_i\geq 1$, $\avgloss_i$ is the average of $b_i$ \iid{}~realizations of $\gloss_i$, which means $\EV[\avgloss_i|b_i]=\gEFi$. Since $\avgloss_i$ and $\avgloss_j$ are independent for $i\neq j$, 
letting the set $B\defn\{b_1,\dots,b_\numDists\}$ we have the identities 
\begin{align}
\EV[\avgloss_i|B] = \EV[\avgloss_i|b_i] = \gEFi 
\quad \text{and} \quad
\EV[\avgloss_i] = \EV[\EV[\avgloss_i|B]] = \EV[\gEFi] = \gEFi. \label{eqn:EVgibar}
\end{align}

\item 
Next, we show that $\EV[b_i/b] = 1/\numDists$. Since all the $b_i$ have \iid{}~statistics, 
$\EV[b_i/b]=\EV[b_j/b] = \mu_1$ for all $i,j\in[\numDists]$, for some constant $\mu_1$. 
We have 
\begin{align}
1 = \EV\left[\sumwi b_i/b\right]=\sumwi \EV[b_i/b] 
= \numDists \mu_1. \label{eqn:EVmu1}
\end{align}
Solving for $\mu_1$ yields the desired result. 
Similarly, note that $\EV[{1/}{b_i}]$ and $\EV[{b_i/}{b^2}]$ are constants independent of $i$, thus we denote them by $\mu_2$ and $\mu_3$ respectively. 

\item 
We denote the variance of a random vector by $\Var$ where $\Var(\cdot) = \EV[\Lnrm{\cdot}^2]-\Lnrm{\EV[\cdot]}^2$. 
The law of total variance states that if $Z$ and $Y$ are random variables on the same probability space, and the variance of $Y$ is finite, then
\[\Var(Y) = \EV[\Var(Y\mid Z)] + \Var(\EV[Y\mid Z]).\]
Using 
the law of total variance and~\eqn{EVgibar}, we upper bound $\Var(\avgloss_i)$ as 
\begin{align}
\Var(\avgloss_i)
= \EV[\Var(\avgloss_i|b_i)] + \Var(\EV[\avgloss_i|b_i])
\leq \EV[\sigma^2/b_i] + \Var(\gEFi) 
\leq \sigma^2\mu_2. \label{eqn:vargi}
\end{align}
The first inequality is due to the assumption $\Var(\gloss_i)\leq\sigma^2$ and the second inequality follows by observing that $\Var(\gEFi)=0$. 
Also, note that from~\eqn{defFw} we have 
\begin{align}
\gEF = \sumworkers \gamma_i\gEFi. \label{eqn:defgFw}
\end{align}
\end{enumerate}

The proofs presented next rely on the the following theorem (\cite{shalev2014understanding} pg.192) that characterizes the convergence of stochastic gradient descent. 
\begin{theorem} \label{thrm:sgdconv}
Let $h:\Rel^d\to\Rel$ be a $\lpzconst$-Lipschitz convex function and $\vvw^* = \argmin_\vvw h(\vvw)$. 
For all $k\in\{0,1,2,\dots\}$ let $v^k$ be a random vector such that $\EV[v^k] = \nabla_\vvw h(\vvw^k)$, and $\Var(v^k) \leq \sigma^2$. 
For a given $\vvw^0$ let the sequence $(\vvw^1, \vvw^2, \dots)$ be such that $\vvw^{k+1} = \vvw^{k} - tv^{k}$. 
Then for a learning rate $t\leq1/\lpzconst$ and  $\bar{\vvw}^k = (\vvw^1 + \dots + \vvw^k)/k$ we have 
$\EV[h(\bar{\vvw}^k)]\leq h(\vvw^*) + \frac{\Ltwo{\vvw^0-\vvw^*}}{2tk} + \frac{t\sigma^2}{2}$.
\end{theorem}
Note that convergence of SGD requires only that $v^k$ be an unbiased estimator of the true gradient and that it have finite variance. 
Next we make use of this observation to prove convergence of the equal and proportional weighting schemes, by showing that they are in fact unbiased estimators of $\gEF$.

\subsubsection{Proof for equal weighting}
From~\eqn{equalweighting2}, all workers posses the same parameter vector and apply the gradient 
$\avglosse \defn \sumwi\gamma_i\avgloss_i$. 
To prove convergence it suffices to show that $\EV[\avglosse]=\gEF$ and that $\varglosse\defn\Var(\avglosse)$ is bounded. 
First note that 
\[\EV[\avglosse] 
= \EV\left[\sumwi\gamma_i\avgloss_i\right] 
= \EV\left[\sumwi\gamma_i\EV[\avgloss_i|b_i]\right] 
= \sumwi\gamma_i\gEFi=\gEF,\] 
which is implied by~\eqn{EVgibar} and~\eqn{defgFw}. Second, using~\eqn{vargi} we have 
\begin{equation}\label{eqn:boundequal}
\varglosse
= \Var\left(\sumwi\gamma_i\avgloss_i\right) 
= \sumwi \Var(\gamma_i\avgloss_i) 
\leq \mu_2\sigma^2\sumwi\gamma_i^2.
\end{equation}
The last equality follows because the $\avgloss_i$ are independent. 
This is the tightest possible bound without making additional assumptions on the distribution of $b_i$. 
Now, according to \thrm{sgdconv}, 
all workers converge to $\wOptGD$.

\subsubsection{Proof for proportional weighting}
All workers update the parameter vectors according to~\eqn{propweighting} using $\avglossp \defn \sumwi \frac{\numDists b_i}{b}\gamma_i\avgloss_i$ as the gradient. 
From \eqn{EVgibar}, \eqn{EVmu1} and \eqn{defgFw} we have 
\[\EV[\avglossp]
= \EV\left[\EV\left[\frac{\numDists}{b}\sumwi b_i\gamma_i\avgloss_i\middle|B\right]\right]
= \EV\left[\frac{\numDists}{b}\sumwi b_i\gamma_i\EV[\avgloss_i|B]\right],\]
which gives 
\[\EV[\avglossp] = \EV\left[\frac{\numDists}{b}\sumwi b_i\gamma_i\gEFi\right]
= \sumwi \gamma_i \gEFi \numDists\EV[b_i/b]
= \gEF.\] 
This result proves $\avglossp$ is an unbiased estimator of $\gEF$. 
To bound $\Var(\avglossp)$ we apply the law of total variance as 
\[\varglossp\defn\Var(\avglossp)
= \underbrace{\EV\left[\Var\left(\frac{\numDists}{b}\sumwi b_i\gamma_i\avgloss_i\middle|B\right)\right]}_{C_1}
+ \underbrace{\Var\left(\EV\left[\frac{\numDists}{b}\sumwi b_i\gamma_i\avgloss_i\middle|B\right]\right)}_{C_2}.\]
Since $b_i\avgloss_i$ is the sum of $b_i$ \iid{}~realizations of $\gloss_i$, 
$\Var(b_i\avgloss_i|B) \leq b_i\sigma^2$. 
This gives us 
\[C_1
\leq \numDists^2\sigma^2\EV\left[\sumwi\frac{b_i}{b^2}\gamma_i^2\right] 
= \numDists^2\mu_3\sigma^2\sumwi\gamma_i^2,\] 
which is the tightest possible bound with the given assumptions on $b_i$.

Next, we bound $C_2$. From~\eqn{EVgibar} we have 
\begin{align}
C_2
=\Var\left(\frac{\numDists}{b}\sumwi \gamma_i b_i\EV[\avgloss_i|B]\right)
=\Var\left(\numDists\sumwi \gamma_i \frac{b_i}{b}\gEFi\right).\label{eqn:c2var}
\end{align}
Noting from~\eqn{EVmu1} that 
\[\EV\left[\numDists\sumwi \gamma_i \frac{b_i}{b}\gEFi\right]
=\numDists\sumwi \gamma_i \EV\left[\frac{b_i}{b}\right]\gEFi
=\sumwi \gamma_i\gEFi
=\gEF,\]
we expand the variance term in~\eqn{c2var} to write 
\[C_2
=\EV\left[\LnrmS{\numDists\sumwi \gamma_i \frac{b_i}{b}\gEFi - \gEF}^2\right]
=\EV\left[\LnrmS{\sumwi \frac{b_i}{b} (\numDists\gamma_i \gEFi - \gEF)}^2\right] 
=\EV\left[\LnrmS{\sumwi \frac{b_i}{b} \Delta_i}^2\right].\] 
Here we use $\Delta_i$ to denote the difference 
$(\numDists\gamma_i \gEFi - \gEF)$. 
Note that from~\eqn{defgFw} we have $\sumwi\Delta_i=0$. 
Letting $c_i\defn\frac{b_i}{b}-\mu_1$ we get $\EV[c_i]=0$ and 
\begin{equation}\label{eqn:definitionofs}
\Var\left(\frac{b_i}{b}\right) = \EV[c_i^2]\defn s^2
\end{equation}
for all $i\in[\numDists]$ for some constant $s$. 
Next we use the identity 
\begin{equation}\label{eqn:identityone}
\sumwi c_i \Delta_i
=\sumwi \left(\frac{b_i}{b}-\mu_1 \right) \Delta_i
=\sumwi \frac{b_i}{b}\Delta_i - \mu_1\sumwi\Delta_i 
=\sumwi \frac{b_i}{b} \Delta_i
\end{equation}
to obtain 
\[C_2
=\EV\left[\LnrmS{\sumwi c_i \Delta_i}^2\right] 
\leq\EV\left[\sumwi c_i^2 \sumwi \Lnrm{\Delta_i}^2\right] 
=\EV\left[\sumwi c_i^2\right] \sumwi \Lnrm{\Delta_i}^2
=\numDists s^2\sumwi \Lnrm{\Delta_i}^2
\leq \numDists^3 s^2 D.\] 
The first equality is due to \eqn{identityone}, and the first inequality is due to triangle inequality and Cauchy-Schwarz inequality. The constant $s$ is as defined in \eqn{definitionofs}.
Note that $\sumwi \Lnrm{\Delta_i}^2$ 
is proportional to the mean squared error of true gradients. 
The last inequality is obtained by assuming that there exists a constant $D\geq0$ such that 
\begin{align}
\sumwi \Lnrm{\Delta_i}^2\leq \numDists^2D. \label{eqn:deltaDbound}
\end{align}
Later we show that in fact such a constant exists.  
Combining the bounds for $C_1$ and $C_2$ give us 
\begin{equation}\label{eqn:boundprop}
\varglossp\leq \numDists^2\mu_3\sigma^2 \sumwi \gamma_i^2 +  \numDists^3s^2 D.
\end{equation}
This result, along with $\EV[\avglossp]=\gEF$ 
prove the convergence of the proportional method.

Now we show that a constant $D\geq0$ exists that satisfies~\eqn{deltaDbound}. 
The $\lpzconst$-Lipschitz continuous assumption on $\loss$ implies $\Lnrm{\nabla_\vvw\loss(\vvw, \rvX)}\leq \lpzconst$. We have 
\[
\Lnrm{\gEFi} 
= \Lnrm{\EV_{\rvX\sim \distQi}[\nabla_\vvw\loss(\vvw, \rvX)]}
\leq \EV_{\rvX\sim \distQi}[\Lnrm{\nabla_\vvw\loss(\vvw, \rvX)}]
\leq \EV[\lpzconst]
= \lpzconst, 
\]
where we get the first inequality by asserting convexity of norms and applying the Jensen's inequality. 
Using this result, an easy albeit loose candidate for $D$ can be obtained as follows. 
We have 
\[\sumwi \Lnrm{\Delta_i}^2
= \sumwi \Lnrm{\numDists\gamma_i \gEFi - \gEF}^2
= \numDists^2 \sumwi \gamma_i^2\Lnrm{\gEFi}^2 - \numDists \Lnrm{\gEF}^2
\leq \numDists^2 \lpzconst^2 \sumwi \gamma_i^2,\]
which lets us take $\lpzconst^2 \sumwi \gamma_i^2$ as $D$. 
This shows the existence of a finite $D$ that satisfies~\eqn{deltaDbound}, and concludes the proof of the convergence of the proportional method.

\subsubsection{Convergence rates comparison}
In this section we compare the two upper bounds obtained for $\varglosse$ and $\varglossp$. 
We start by showing that $\numDists^2\mu_3\leq \mu_2$. We have 
$\EV[b_i|b] = \EV[b_j|b] = \frac{1}{\numDists}\sumwi\EV[b_i|b]
= \frac{1}{\numDists}\EV[\sumwi b_i|b] = \frac{b}{\numDists}$. 
Since the reciprocal function is convex, by Jensen's inequality 
$\EV\left[\frac{1}{b_i}\middle|b\right]\geq \frac{1}{\EV[b_i|b]} = \frac{\numDists}{b}$. 
Now we have 
\[\EV\left[\frac{\numDists}{b}-\frac{1}{b_i}\right]
=\EV\left[\frac{\numDists}{b}-\EV\left[\frac{1}{b_i}\middle|b\right]\right]
\leq\EV\left[\frac{\numDists}{b}- \frac{\numDists}{b}\right]
=0,\] 
which gives 
$\EV\left[\frac{\numDists}{b}\right] \leq \EV\left[\frac{1}{b_i}\right]$. 
Therefore, 
\begin{align}
\numDists^2\mu_3
=\numDists^2\EV\left[\frac{b_i}{b^2}\right]
=\numDists^2\EV\left[\EV\left[\frac{b_i}{b^2}\middle|b\right]\right]
=\EV\left[\frac{\numDists^2}{b^2}\EV\left[b_i|b\right]\right],
\end{align}
which gives
\begin{align}
\numDists^2\mu_3
=\EV\left[\frac{\numDists^2}{b^2}\EV\left[b_i|b\right]\right]
=\EV\left[\frac{\numDists^2}{b^2}\frac{b}{\numDists}\right]
=\EV\left[\frac{\numDists}{b}\right]
\leq \EV\left[\frac{1}{b_i}\right]
= \mu_2.   \label{eqn:nsquaredmu3mu2}
\end{align}

As suggested by the SGD convergence properties summarized in \thrm{sgdconv}, the proportional method converges faster if $\varglossp \leq \varglosse$. 
The two upper bounds we obtained for $\varglossp$ and $\varglosse$ are the tightest possible bounds without making additional assumptions on $b_i$. Therefore, in general we expect the proportional method to converge faster if the upper bound for $\varglossp$ is less than that for $\varglosse$. 
Substituting from \eqn{boundequal} and \eqn{boundprop} to the upper bounds we have
\[\numDists^2\mu_3\sigma^2 \sumwi \gamma_i^2 +  \numDists^3s^2 D 
\leq \mu_2\sigma^2\sumwi\gamma_i^2.\]
Rearranging the terms gives the condition for the proportional method to converge faster than equal weighting
\begin{equation}\label{eqn:icasspcondition}
{D/\sigma^2} \leq (\mu_2 - \numDists^2\mu_3)\sumwi \gamma_i^2/(\numDists^3s^2).
\end{equation}
The right size is non-negative due to the result in \eqn{nsquaredmu3mu2}.

This condition is insightful. The right side depends only on the straggler statistics.
Recall that $\sigma^2$ and $D$ are due to our assumptions $\Var(\gloss_i)\leq\sigma^2$ and 
$\sumwi \Lnrm{\Delta_i}^2 \leq \numDists^2D$. 
While $D$ measures how different the true gradients across workers are (global variation), $\sigma^2$ measures the variance of the gradient distribution within one worker (local variance). 
The inequality binds together attributes of three different phenomenons: stragglers, measurement noise and data distributions. 
If the other factors remain the same, a large $\sigma^2$ is likely to satisfy the condition, making the proportional method converge faster.

\begin{figure}
\centering
\includegraphics[width=0.88\textwidth]{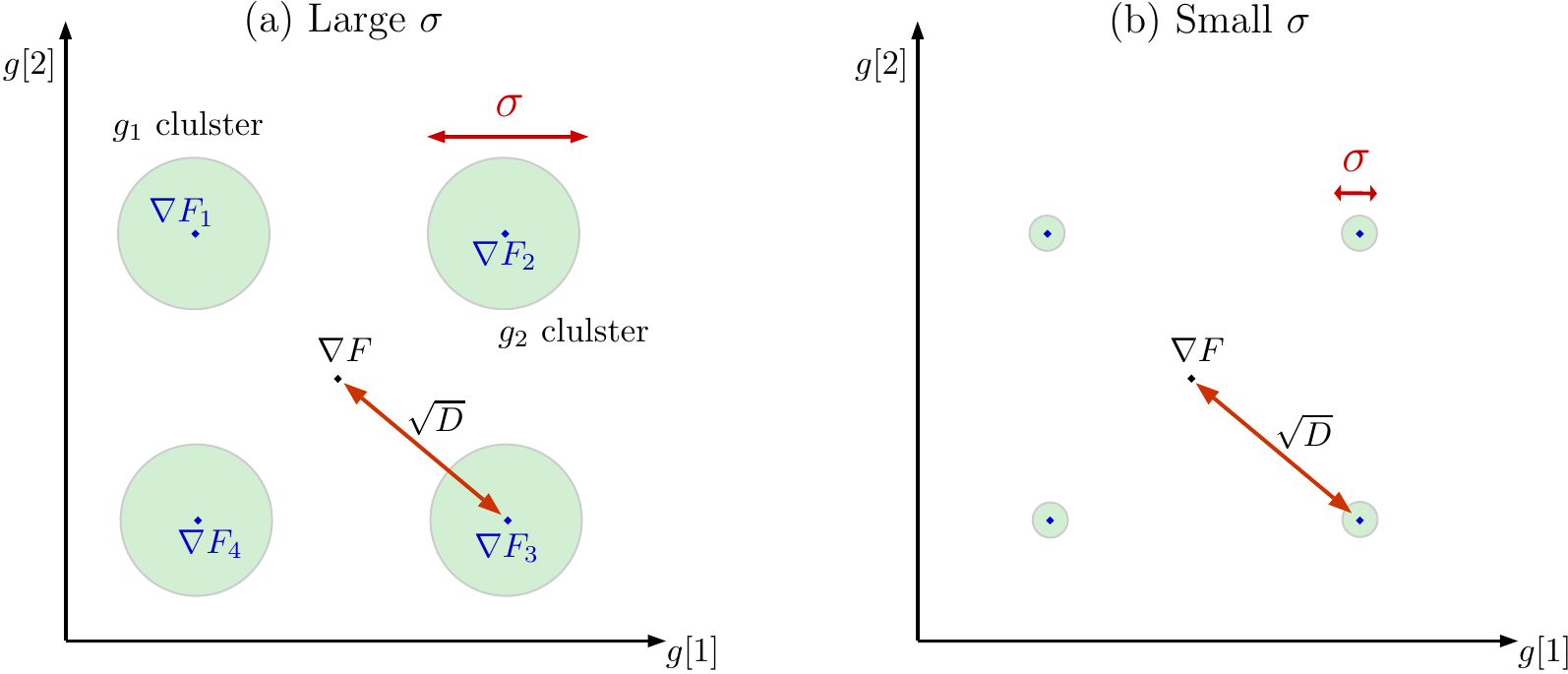}
\caption{Visualizing \eqn{icasspcondition}, the condition for the proportional method to converge faster than equal weighting.}
\label{fig:icassp_visualizing_condition}
\end{figure}

In \fig{icassp_visualizing_condition} we attempt to visualize the condition in \eqn{icasspcondition} using an example. 
In this example we assume a 2-dimensional gradient. The $x$ and $y$ axes represent the two components of the gradient. 
Let $\gamma_i=1/4$, and we assume that $\vvw$ is fixed. 
In the left figure, the four shaded clusters represent distributions of the random variables $\nabla_\vvw\loss(\vvw,\rvX); \rvX\sim\distQi$ for $i\in[4]$. At the centre of each cluster is the expected gradient $\gEFiw = \gEFi = \EV_{\rvX\sim \distQi}[\nabla_\vvw\loss(\vvw, \rvX)]$. 
The variance of $\nabla_\vvw\loss(\vvw,\rvX)$ is $\sigma^2$ in all clusters. As per~\eqn{defgFw} 
$\gEF = \frac{1}{4}\sum_{i=0}^{4}\gEFi$ and per \eqn{deltaDbound} $\gEF$ has a distance $\sqrt{D}$ to all cluster centres. 
The right figure is same as the left, except that its $\sigma^2$ is smaller. 
In both figures $D$ is the same. 
The figures illustrate that $\sigma^2$ is a measurement local to the workers (a single shaded region), and $D$ is a measure of the global variation among workers (all four shaded regions). 
Going back to~\eqn{icasspcondition}, we observe that a smaller $\sigma^2$ is likely to violate the inequality, making the equal method converge faster. 
As per the figure on the right, a smaller $\sigma^2$ also implies a smaller noise radius. 
Even $b_i=1$ is enough to estimate a cluster centre $\gEFi$ with a relatively high accuracy, and the error will be low compared to $D$. 
However, if $\sigma^2$ is large, we need a large $b_i$ to keep the measurement error low. 
In such cases the proportional method performs better by leveraging the gradients with higher confidence. 

Note that $D$ is an implicit measure of the divergence of worker distributions $\distQ_i$, which is observed through $\gEFi$s. 
Authors of~\cite{ferdinand2018anytime} consider the case when distributions of workers are the same. 
This means we have 
$\gEFi=\gEF$, and we are allowed to set $D=0$ to get the tightest bound in~\eqn{deltaDbound}. 
If $D=0$, \eqn{nsquaredmu3mu2} implies that the condition in~\eqn{icasspcondition} is always true. 
In this case the proportional scheme is guaranteed to perform better, and is inline with the findings in \cite{ferdinand2018anytime}. 
However, when $D\neq0$ the proportional scheme outperforms equal weighting only if $D/\sigma^2$ is small enough. 
This means no matter how noisy the $\avgloss_i$ (measured through $\sigma^2$) from some workers are, they cannot be weighted down if the gradient distributions are significantly different (observed through $D$). 
While $D/\sigma^2$ is not measurable for real world datasets, 
the proportional weighting method performs better in the experiments we present in \secn{icasspexperiments}.
Therefore, the condition \eqn{icasspcondition} seems to be satisfied in those experiments.

\section{Optimal gradient estimator} \label{secn:icasspoptimality}
In \secn{heuristicsec} we analyzed two methods of combining the gradient estimates of workers. We showed that $\avglosse$ and $\avglossp$ corresponding to the equal and proportional methods are unbiased estimators of $\gEF$ and have finite variances. As per \thrm{sgdconv}, the unbiased estimator with the smallest variance wins the gradient descent race. We would like to know if there exist any other estimators that achieve a smaller variance. 
In this section we make use of the theory on `minimum variance unbiased estimator' (MVUE) to answer the question.

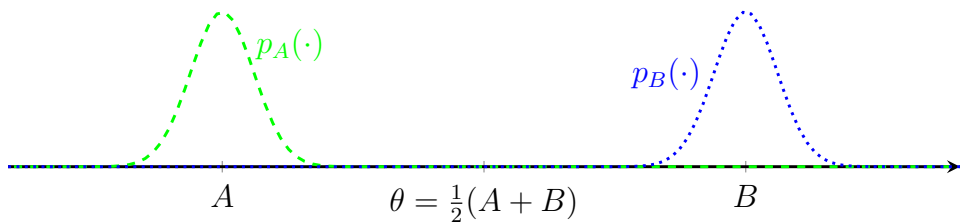
\begin{figure}
	\centering\begin{tikzpicture}
\begin{axis}[
xmin=-5, xmax=5,
axis y line=none,
axis x line=middle,
line width=1.1pt,
width=14cm,
height=4cm,
xtick={-3.3, 0, 3.3},
xticklabels={$A$, $\theta = \frac{1}{2}(A+B)$, $B$},
enlargelimits]
\addplot[green, dashed, domain=-10:10, samples=200]  {pow(4,-((x+3.3)*1.5)^2)} node[right, pos=0.359]{$p_A(\cdot)$};
\addplot[blue, dotted, domain=-10:10, samples=200] {pow(4,-((x-3.3)*1.5)^2)} node[left, pos=0.63] {$p_B(\cdot)$};
\end{axis}
\end{tikzpicture}
	\caption{Samples are drawn from $p_A$ and $p_B$ with the goal of estimating $\theta$.}\label{fig:exampleTwoWorkersAB}
\end{figure}

To better understand the problem at hand, we start with the toy problem  illustrated in \fig{exampleTwoWorkersAB}, which is obtained by assuming two workers and a scalar parameter vector. We assume the parameter vector is fixed and $\gamma_1=\gamma_2=\frac{1}{2}$. 
Let $\gEF_1=A$ and $\gEF_2=B$ be the true gradients at the first and second workers respectively. 
Our goal is to obtain the minimum variance estimate of $\theta = \frac{1}{2}(A+B)$, the average gradient of the two workers. 
We assume the gradients sampled at the two workers are normally distributed with variance $\sigma^2$, and means $A$ and $B$. This distribution is due to the randomness of data available at each worker. 
The probability distributions for the samples drawn at the workers are denoted by $p_A$ and $p_B$. 
Let the first worker obtain $a_1, \dots, a_M$ 
\iid{} samples and the second worker to obtain $b_1, \dots, b_N$ 
\iid{} samples. 
Now we apply \thrm{crlb} on the Cramer-Rao Lower Bound (CRLB) to this problem. 

\begin{theorem}[Cramer-Rao Lower Bound - Scalar parameter~\cite{kay1993fundamentals}] \label{thrm:crlb}
	For an observation vector $\vvbx$ it is assumed that the PDF $p(\vvbx; \theta)$ satisfies the `regularity' condition $\EV[\frac{\partial p(\vvbx; \theta)}{\partial\theta}] = 0$ for all $\theta$, where the expectation is taken with respect to $p(\vvbx; \theta)$. Then, the variance of any unbiased estimator $\hat{\theta}$ must satisfy $\Var(\hat{\theta}) \geq - \EV[\frac{\partial^2 p(\vvbx; \theta)}{\partial\theta^2}]^{-1}$,
	where the derivative is evaluated at the true value of $\theta$ and the expectation is taken with respect to $p(\vvbx; \theta)$. Furthermore, an unbiased estimator may be found that attains the bound for all $\theta$ if and only if $\frac{\partial p(\vvbx; \theta)}{\partial\theta} = I(\theta)(g(\vvbx)-\theta)$ for some functions $g$ and $I$. That estimator which is the MVUE is $\hat{\theta} = g(x)$, and the minimum variance is $1/I(\theta)$. 
\end{theorem}

We look at the following two cases and apply CRLB.

\subsection{Constant observation counts} \label{secn:mnconst}
Let $M$ and $N$ be constants and let the observation vector be defined as 
\[\vvx = [a_1, \dots, a_M, b_1, \dots, b_N].\] 
Since all samples are independent we have the joint PDF
\[p = p(\vvx; \theta)
=\prod_{m=1}^{M} p_A(a_m) \prod_{n=1}^{N} p_B(b_n)
=\prod_{m=1}^{M} \frac{1}{\sqrt{2\pi\sigma^2}} e^{-\frac{(a_m-A)^2}{2\sigma^2}}
\prod_{n=1}^{N} \frac{1}{\sqrt{2\pi\sigma^2}} e^{-\frac{(b_n-B)^2}{2\sigma^2}}.\]
By taking partial derivative of $\ln p$ with respective to $\theta$ we have 
\begin{align} 
\dPdT &= \frac{1}{\sigma^2} \sum_{m=1}^{M}(a_m-A)\dAdT + \frac{1}{\sigma^2}\sum_{n=1}^{N}(b_n-B)\dBdT.\label{eqn:dabalnp}
\end{align}
Since
\[\EV\left[\dPdT\right] = \frac{1}{\sigma^2} \sum_{m=1}^{M}(\EV[a_m]-A)\dAdT + \frac{1}{\sigma^2}\sum_{n=1}^{N}(\EV[b_n]-B)\dBdT = 0,\]
the PDF satisfies the regularity condition and we can apply CRLB. 
Also note that $\theta = \frac{1}{2}(A+B)$ we have $\dTdA = \dTdB = \frac{1}{2}$ which gives $\dAdT = \dBdT = 2$. 
Differentiating \eqn{dabalnp} with respective to $\theta$ gives 
\[\dtPdTt = \frac{1}{\sigma^2} \sum_{m=1}^{M}\left((a_m-A)\dtAdTt - \left(\dAdT\right)^2\right) 
+ \frac{1}{\sigma^2} \sum_{n=1}^{N}\left((b_n-B)\dtBdTt - \left(\dBdT\right)^2\right)\]
and by taking expectation and substituting $\dAdT = \dBdT = 2$ we have 
$\EV\left[\dtPdTt\right] = -\frac{4(M+N)}{\sigma^2}$.
This means for any unbiased estimator $\hat{\theta}$, by CRLB $\Var({\hat{\theta}}) \geq \frac{\sigma^2}{4(M+N)}$. 
To find an estimator that achieves the minimum variance we rewrite \eqn{dabalnp} as follows:
\begin{align} 
\dPdT
&=\frac{2}{\sigma^2} \sum_{m=1}^{M}(a_m-A) + \frac{2}{\sigma^2}\sum_{n=1}^{N}(b_n-B) \nonumber \\
&=\frac{2}{\sigma^2} \left(\sum_{m=1}^{M}a_m + \sum_{n=1}^{N}b_n - (MA+NB)\right). \label{eqn:mvueexist}
\end{align}
Note that \eqn{mvueexist} cannot be written in $I(\theta)(g(\vvx)-\theta)$ form for $\theta = \frac{1}{2}(A+B)$ and we conclude that the MVUE does not exist.

Knowing that an estimator achieving CRLB does not exist, we can compare the variances of the two estimators discussed in \secn{heuristicsec} with the lower bound. 
Gradient estimator for the equal method is given by $\avglosse = \frac{1}{2}\left(\frac{1}{M}\sum_{m=1}^{M} a_m + \frac{1}{N}\sum_{n=1}^{N} b_n \right)$. We can verify $\avglosse$ is in fact an unbiased estimator by noting that $\EV[\avglosse] = \frac{1}{2}(A+B) = \theta$. For variance we have
\begin{align*}
\Var(\avglosse)
&= \frac{1}{4}\left(\frac{1}{M^2}\sum_{m=1}^{M} \Var(a_m) + \frac{1}{N^2}\sum_{n=1}^{N} \Var(b_n) \right)\\
&= \frac{\sigma^2}{4}\left(\frac{1}{M} + \frac{1}{N} \right)\\
&= \frac{\sigma^2}{4 (M+N)}\frac{(M+N)^2}{MN}\\
&= \frac{\sigma^2}{4 (M+N)}\left(\frac{M}{N} + \frac{N}{M} + 2\right)\\
&\geq \frac{\sigma^2}{(M+N)},
\end{align*}
where we used the fact that the observations are independent, and $\frac{M}{N} + \frac{N}{M}\geq 2$ for $M,N\geq 1$. 
This shows that the variance of equal estimator $\Var(\avglosse)$ is at least $4$ times larger that the lower bound. 
For the proportional method the gradient estimator is 
\[\avglossp = \frac{M}{M+N}\left(\frac{1}{M}\sum_{m=1}^{M} a_m\right) + \frac{N}{M+N}\left(\frac{1}{N}\sum_{n=1}^{N} b_n \right)
= \frac{1}{M+N}\left(\sum_{m=1}^{M} a_m + \sum_{n=1}^{N} b_n \right)
.\] 
In this case $\avglosse$ is not an unbiased estimator of $\theta$ since $\EV[\avglossp] = \frac{MA+NB}{M+N} \neq \theta$. However for variance we have
\[\Var(\avglossp)
= \frac{1}{(M+N)^2}(M{\sigma^2} + N\sigma^2)
= \frac{\sigma^2}{M+N},\]
which is again larger than the CRLB. Next we assume $M$ and $N$ are \iid{} RVs and redo the analysis. 

\subsection{Random observation counts}
Let $M$ and $N$ be \iid{} random variables. 
In this case the observation vector itself is of random length. Let $q(\cdot)$ be the PDF of $M$ and $N$. We assume $\EV[M] = \EV[N] = \mu$, and $\EV[1/M] = \EV[1/N] = \mu_2$. 
We assume that $q$ is \emph{not} a function of $\theta$. This assumption is consistent with our distributed optimization setup where the $b_i$ are independent of $\gEF_i$.
Now we have $p = p(\vvx \mid M,N ; \theta)p_{M,N}(M,N)$. The first term is same as what we derived in \secn{mnconst}. By observing $M$ and $N$ are independent we write 
\[\ln p = \ln p(\vvx \mid M,N ; \theta)  + \ln q(M) + \ln q(N).\] 
Noting that $\theta$ is not a parameter in $q$, we differentiate with respect to $\theta$ to obtain \eqn{dabalnp} once again. 
Therefore, $\EV\left[\dPdT\right] = 0$ for this case as well. 
The CRLB in this case is given by 
\[\EV\left[\dtPdTt\right] = \EV\left[-\frac{4(M+N)}{\sigma^2} \right] = -\frac{4}{\sigma^2}\EV[M+N] = -\frac{8\mu}{\sigma^2},\]
which means for any unbiased estimator $\hat{\theta}$, $\Var({\hat{\theta}}) \geq \frac{\sigma^2}{8\mu}$. 
Regarding the existence of an estimator achieving the lower bound, the same argument as before applies. We cannot write \eqn{mvueexist} in  $I(\theta)(g(\vvx)-\theta)$ form, and MVUE does not exist. 

Let us now compare the lower bound with the variances of $\avglosse$ and $\avglossp$. The estimator $\avglosse$ remains unbiased and we have 
\[\Var(\avglosse) = \frac{1}{4}\left(\Var\left(\frac{1}{M}\sum_{m=1}^{M} a_m\right) + \Var\left(\frac{1}{N}\sum_{n=1}^{N} b_n\right) \right)\]
and for the first (and second) variance term
\begin{align*}
\Var\left(\frac{1}{M}\sum_{m=1}^{M} a_m\right)
&= \EV\left[\Var\left( \frac{1}{M}\sum_{m=1}^{M} a_m \middle| M \right) \right] 
+ \Var\left(\EV\left[ \frac{1}{M}\sum_{m=1}^{M} a_m \middle| M\right ] \right) \\
&= \EV\left[\frac{1}{M^2}\Var\left(\sum_{m=1}^{M} a_m \middle| M \right) \right] 
+ \Var\left(\EV[A \middle| M] \right)\\
&= \EV\left[\frac{1}{M^2}\sigma^2 M \right] + 0 \\
&= \sigma^2 \mu_2.
\end{align*}
Substituting the result we get $\Var(\avglosse) = \frac{\sigma^2 \mu_2}{2}$. 
The reciprocal of a positive real number is a convex function and by Jensen's inequality we can show that $\frac{1}{\mu} = \frac{1}{\EV[M]} \leq \EV[\frac{1}{M}] = \mu_2$. We again conclude that $\Var(\avglosse)$ is at least $4$ times larger than the CRLB. 
In contrast to the \secn{mnconst} proportional estimator now is unbiased since
\begin{align*}
\EV[\avglossp] 
&= \EV\left[\frac{M}{M+N}\left(\frac{1}{M}\sum_{m=1}^{M} a_m\right) 
+ \frac{N}{M+N}\left(\frac{1}{N}\sum_{n=1}^{N} b_n \right)\right]\\
&= \EV\left[\frac{M}{M+N}\EV\left[\frac{1}{M}\sum_{m=1}^{M} a_m \middle| M,N \right]\right] 
+ \EV\left[\frac{N}{M+N}\EV\left[\frac{1}{N}\sum_{n=1}^{N} b_n \middle| M,N \right]\right]\\
&= \EV\left[\frac{M}{M+N}\right] A 
+ \EV\left[\frac{N}{M+N}\right]B\\
&= \frac{1}{2}(A+B).
\end{align*}
The last equality is because $\EV\left[\frac{M}{M+N}\right] = \EV\left[\frac{N}{M+N}\right] = \frac{1}{2}$ for \iid{} $M$, $N$. For the variance we can show that $\Var(\avglossp)\geq \sigma^2 \EV\left[\frac{1}{M+N}\right]$ which is greater than the CRLB. Therefore, we can conclude that neither of the two methods achieve the minimum variance, and, more importantly no other estimator does. 

\section{Conclusions and future work} \label{secn:future}
In this paper we study the convergence of decentralized optimization when data distributions at workers are non-identical \emph{and} when workers perform variable amounts of work. 
We numerically and theoretically analyze two heuristic methods of combining worker outputs. 
We make use of the theory on minimum variance unbiased estimator (MVUE) to evaluate the existence of an optimal method for combining worker outputs. 
While we conclude that neither of the two heuristic methods are optimal, we also show that an optimal method does \emph{not} exist. 
A few possible next steps to this theme of works is as follows. 

First is improving the theoretical analysis presented in \secn{analysis}. We provide a convergence proof for the proportional method when consensus is perfect. We would like to extend the convergence proof and generalize the condition in~\eqn{icasspcondition} to approximate consensus. Such an analysis will largely benefit from the work in \cite{duchi2011dual} as there are many similarities between the algorithms. 
Also, in our analysis we assume that $b_i\geq 1$. A natural next step will be to allow $b_i = 0$ and generalize our convergence results accordingly. This inclusion means that some workers may not produce a gradient estimate at all, therefore, may skip over some iterations. In other words, only a subset of workers participate in each iteration \cite{konevcny2016federatedlearning}. 

Second next step is regarding the interplay between worker connectivity graph and worker distributions. In our work we consider that data distributions at workers are different. However, in reality it is safe to assume that the distributions are similar to some extent, but with a few workers having `atypical' distributions scattered around. It will be interesting to understand how the position on the graph of these atypical workers impact the convergence. For example, assume three workers whose network connectivity makes a chain. We have two workers with similar distributions and one worker with a different distribution. It will be interesting to see whether a faster convergence is obtained by placing the outlier distribution in middle of the chain or at one end of the chain.
We leave the consideration of this type of a system model as future work. 

\section*{Acknowledgment}
The authors would like to thank Haider Al-Lawati at University of Toronto, Jason Lam and Zhenhua Hu at Huawei Technologies Canada for technical discussions, 
and Compute Canada (\href{http://www.computecanada.ca}{www.computecanada.ca}) for providing computing resources for the experiments.



\end{document}